\documentclass[10pt]{article}
\usepackage{amsfonts}

\usepackage{amsmath}
\usepackage{epsfig}

\begin{document}

\title{\textbf{The fractional volatility model: An agent-based interpretation}}
\author{\textbf{R. Vilela Mendes}\thanks{%
Centro de Matem\'{a}tica e Aplica\c{c}\~{o}es Fundamentais and Universidade
T\'{e}cnica de Lisboa, e-mail: vilela@cii.fc.ul.pt} \\
Complexo Interdisciplinar,\\
Av. Prof. Gama Pinto 2, 1649-003 Lisboa, Portugal}
\date{}
\maketitle

\begin{abstract}
Based on criteria of mathematical simplicity and consistency with empirical
market data, a model with volatility driven by fractional noise has been
constructed which provides a fairly accurate mathematical parametrization of
the data. Here, some features of the model are discussed and, using
agent-based models, one tries to find which agent strategies and (or)
properties of the financial institutions might be responsible for the
features of the fractional volatility model.
\end{abstract}


\section{Introduction}

Classical Mathematical Finance has, for a long time, been based on the
assumption that the price process of market securities may be approximated
by geometric Brownian motion 
\begin{equation}
\begin{array}{lll}
dS_{t} & = & \mu S_{t}dt+\sigma S_{t}dB\left( t\right)
\end{array}
\label{1.00}
\end{equation}
In liquid markets the autocorrelation of price changes decays to negligible
values in a few minutes, consistent with the absence of long term
statistical arbitrage. Geometric Brownian motion models this lack of memory,
although it does not reproduce the empirical leptokurtosis. On the other
hand, nonlinear functions of the returns exhibit significant positive
autocorrelation. For example, there is volatility clustering, with large
returns expected to be followed by large returns and small returns by small
returns (of either sign). This, together with the fact that autocorrelations
of volatility measures decline very slowly\cite{Ding2} \cite{Harvey} \cite
{Crato}, has the clear implication that long memory effects should somehow
be represented in the process and this is not included in the geometric
Brownian motion hypothesis.

One other hand, as pointed out by Engle\cite{Engle}, when the future is
uncertain, investors are less likely to invest. Therefore uncertainty
(volatility) would have to be changing over time. The conclusion is that a
dynamical model for volatility is needed and $\sigma $ in Eq.(\ref{1.00}),
rather than being a constant, becomes a process by itself. This idea led to
many deterministic and stochastic models for the volatility (\cite{Taylor} 
\cite{Engle2} and references therein).

In a previous paper\cite{VilOliv}, using both a criteria of mathematical
simplicity and consistency with market data, a stochastic volatility model
was constructed, with volatility driven by fractional noise. It appears to
be the minimal model consistent both with mathematical simplicity and the
market data. This data-inspired model is different from the many stochastic
volatility models that have been proposed in the literature. The model was
used to compute the price return statistics and asymptotic behavior, which
were compared with actual data. Deviations from the classical Black-Scholes
result and a new option pricing formula were also obtained. The \textit{%
fractional volatility model}, its predictions and comparison with data will
be reviewed in Section 2.

When this fractional volatility model was first presented, an interesting
remark by an economist was ''\textit{all right, the model seems to fit
reasonably well the data, but where is the economics ? }''. The same remark
might be made about the simple geometric Brownian model, which does not even
fit the data and is used by most of the of mathematical finance
practitioners.. But, of course, our economist was right. The fractional
volatility model seems to be a reasonable \textit{mathematical
parametrization} of the market behavior, but it is not sufficient to fit the
data. One should also search for the mechanisms in the market that lead to
the observed phenomena. No agent-based model can pretend to be the market
itself, not even a realistic image of it. Nevertheless it may provide a
surrogate model of the basic mechanics at work there. Therefore, the idea in
this paper is to use stylized agent-based market models and find out which
features of these models correspond to each one of the features of the
mathematical parametrization of the data.

\section{The fractional volatility model. Induced volatility, statistics of
returns, option pricing and leverage}

The basic hypothesis for the model construction were:

(H1) The log-price process $\log S_{t}$ belongs to a probability product
space $\Omega \otimes \Omega ^{^{\prime }}$ of which the first one, $\Omega $%
, is the Wiener space and the second, $\Omega ^{^{\prime }}$, is a
probability space to be reconstructed from the data. Denote by $\omega \in
\Omega $ and $\omega ^{^{\prime }}\in \Omega ^{^{\prime }}$ the elements
(sample paths) in $\Omega $ and $\Omega ^{^{\prime }}$ and by $\mathcal{F}%
_{t}$ and $\mathcal{F}_{t}^{^{\prime }}$ the $\sigma -$algebras in $\Omega $
and $\Omega ^{^{\prime }}$ generated by the processes up to $t$. Then, a
particular realization of the log-price process is denoted 
\[
\log S_{t}\left( \omega ,\omega ^{^{\prime }}\right) 
\]
This first hypothesis is really not limitative. Even if none of the
non-trivial stochastic features of the log-price were to be captured by
Brownian motion, that would simply mean that $S_{t}$ is a trivial function
in $\Omega $.

(H2) The second hypothesis is stronger, although natural. One assumes that,
for each fixed $\omega ^{^{\prime }}$, $\log S_{t}\left( \bullet ,\omega
^{^{\prime }}\right) $ is a square integrable random variable in $\Omega $.

A mathematical consequence of hypothesis (H2) is that, for each fixed $%
\omega ^{^{\prime }}$, 
\begin{equation}
\begin{array}{lll}
\frac{dS_{t}}{S_{t}}\left( \bullet ,\omega ^{^{\prime }}\right) & = & \mu
_{t}\left( \bullet ,\omega ^{^{\prime }}\right) dt+\sigma _{t}\left( \bullet
,\omega ^{^{\prime }}\right) dB\left( t\right)
\end{array}
\label{2.1}
\end{equation}
where $\mu _{t}\left( \bullet ,\omega ^{^{\prime }}\right) $ and $\sigma
_{t}\left( \bullet ,\omega ^{^{\prime }}\right) $ are well-defined processes
in $\Omega $. (Theorem 1.1.3 in Ref.\cite{Nualart})

Recall that if $\left\{ X_{t},\mathcal{F}_{t}\right\} $ is a process such
that 
\begin{equation}
\begin{array}{lll}
dX_{t} & = & \mu _{t}dt+\sigma _{t}dB\left( t\right)
\end{array}
\label{2.2}
\end{equation}
with $\mu _{t}$ and $\sigma _{t}$ being $\mathcal{F}_{t}-$adapted processes,
then 
\begin{equation}
\begin{array}{lll}
\mu _{t} & = & \underset{\varepsilon \rightarrow 0}{\lim }\frac{1}{%
\varepsilon }\left\{ \left. E\left( X_{t+\varepsilon }-X_{t}\right) \right| 
\mathcal{F}_{t}\right\} \\ 
\sigma _{t}^{2} & = & \underset{\varepsilon \rightarrow 0}{\lim }\frac{1}{%
\varepsilon }\left\{ \left. E\left( X_{t+\varepsilon }-X_{t}\right)
^{2}\right| \mathcal{F}_{t}\right\}
\end{array}
\label{2.3}
\end{equation}

The process associated to the probability space $\Omega ^{^{\prime }}$ could
then be inferred from the data. According to (\ref{2.3}), for each fixed $%
\omega ^{^{\prime }}$ realization in $\Omega ^{^{\prime }}$ one has 
\begin{equation}
\sigma _{t}^{2}\left( \bullet ,\omega ^{^{\prime }}\right) =\underset{%
\varepsilon \rightarrow 0}{\lim }\frac{1}{\varepsilon }\left\{ E\left( \log
S_{t+\varepsilon }-\log S_{t}\right) ^{2}\right\}  \label{2.4}
\end{equation}
Each set of market data corresponds to a particular realization $\omega
^{^{\prime }}$. Therefore, assuming the realization to be typical, the $%
\sigma _{t}^{2}$ process may be reconstructed from the data by the use of (%
\ref{2.4}). This data-reconstructed $\sigma _{t}$ process was called the 
\textit{induced volatility}.

For practical purposes we cannot strictly use Eq.(\ref{2.4}) to reconstruct
the induced volatility process, because when the time interval $\varepsilon $
is very small the empirical evaluation of the variance becomes unreliable.
Instead, $\sigma _{t}$ was estimated from 
\begin{equation}
\sigma _{t}^{2}=\frac{1}{\left| T_{0}-T_{1}\right| }\mathnormal{var}\left(
\log S_{t}\right)  \label{2.5}
\end{equation}
with a time window $\left| T_{0}-T_{1}\right| $ sufficiently small to give a
reasonably local characterization of the volatility, but also sufficiently
large to allow for a reliable estimate of the local variance of $\log S_{t}$.

Once several data sets were analyzed\cite{VilOliv}, the next step towards
obtaining a mathematical characterization of the \textit{induced volatility
process} was to look for scaling properties. It turned out that neither 
\begin{equation}
E\left| \sigma \left( t+\Delta \right) -\sigma \left( t\right) \right| \sim
\Delta ^{H}  \label{2.6}
\end{equation}
nor 
\begin{equation}
E\left| \frac{\sigma \left( t+\Delta \right) -\sigma \left( t\right) }{%
\sigma \left( t\right) }\right| \sim \Delta ^{H}  \label{2.7}
\end{equation}
were good hypothesis for the induced volatility process. It means that the
induced volatility process itself is not self-similar.

If instead, one computes the empirical integrated log-volatility, one finds
that it is well represented by a relation of the form 
\begin{equation}
\sum_{n=0}^{t/\delta }\log \sigma \left( n\delta \right) =\beta t+R_{\sigma
}\left( t\right)  \label{2.8}
\end{equation}
the $R_{\sigma }\left( t\right) $ process possessing very accurate
self-similar properties.

A nondegenerate process $X_{t}$, if it has finite variance, stationary
increments and is self-similar 
\begin{equation}
\mathnormal{Law}\left( X_{at}\right) =\mathnormal{Law}\left(
a^{H}X_{t}\right)  \label{2.9}
\end{equation}
must necessarily \cite{Embrechts} have a covariance 
\begin{equation}
\mathnormal{Cov}\left( X_{s},X_{t}\right) =\frac{1}{2}\left( \left| s\right|
^{2H}+\left| t\right| ^{2H}-\left| s-t\right| ^{2H}\right) E\left(
X_{1}^{2}\right)  \label{2.10}
\end{equation}
with $0<H\leq 1$. The simplest process with these properties is a Gaussian
process called fractional Brownian motion\cite{Mandelbrot1}, with 
\begin{equation}
\mathbb{E}\left[ B_{H}\left( t\right) \right] =0\qquad \mathbb{E}\left[
B_{H}\left( t\right) B_{H}\left( s\right) \right] =\frac{1}{2}\left\{ \left|
t\right| ^{2H}+\left| s\right| ^{2H}-\left| t-s\right| ^{2H}\right\}
\label{2.11}
\end{equation}
and, for $H>\frac{1}{2}$ , a long range dependence 
\begin{equation}
\sum_{n=1}^{\infty }\mathnormal{Cov}\left( B_{H}\left( 1\right) ,B_{H}\left(
n+1\right) -B_{H}\left( n\right) \right) =\infty  \label{2.12}
\end{equation}

Therefore, mathematical simplicity suggested the identification of the $%
R_{\sigma }\left( t\right) $ process with fractional Brownian motion. 
\begin{equation}
R_{\sigma }\left( t\right) =kB_{H}\left( t\right)  \label{2.13}
\end{equation}
and, from the data, one obtains Hurst coefficients in the range $0.8-0.9$.

Finally one obtains the following \textit{fractional volatility model}

\begin{equation}
\begin{array}{lll}
dS_{t} & = & \mu S_{t}dt+\sigma _{t}S_{t}dB\left( t\right) \\ 
\log \sigma _{t} & = & \beta +\frac{k}{\delta }\left\{ B_{H}\left( t\right)
-B_{H}\left( t-\delta \right) \right\}
\end{array}
\label{2.16}
\end{equation}
$k$ is a volatility intensity parameter and $\delta $ is the observation
time scale. Notice that the volatility is not driven by fractional Brownian
motion but by fractional noise, naturally introducing an observation scale
dependence.

\subsection{The statistics of price returns}

At each fixed time $\log \sigma _{t}$ is a Gaussian random variable with
mean $\beta $ and variance $k^{2}\delta ^{2H-2}$. Then, 
\begin{equation}
p_{\delta }\left( \sigma \right) =\frac{1}{\sigma }p_{\delta }\left( \log
\sigma \right) =\frac{1}{\sqrt{2\pi }\sigma k\delta ^{H-1}}\exp \left\{ -%
\frac{\left( \log \sigma -\beta \right) ^{2}}{2k^{2}\delta ^{2H-2}}\right\}
\label{3.2}
\end{equation}
therefore

\begin{equation}
P_{\delta }\left( \log \frac{S_{T}}{S_{t}}\right) =\int_{0}^{\infty }d\sigma
p_{\delta }\left( \sigma \right) p_{\sigma }\left( \log \frac{S_{T}}{S_{t}}%
\right)  \label{3.3}
\end{equation}
with 
\begin{equation}
p_{\sigma }\left( \log \frac{S_{T}}{S_{t}}\right) =\frac{1}{\sqrt{2\pi
\sigma ^{2}\left( T-t\right) }}\exp \left\{ -\frac{\left( \log \left( \frac{%
S_{T}}{S_{t}}\right) -\left( \mu -\frac{\sigma ^{2}}{2}\right) \left(
T-t\right) \right) ^{2}}{2\sigma ^{2}\left( T-t\right) }\right\}  \label{3.4}
\end{equation}
Thus, the effective probability distribution of the returns might depend
both on the time lag $\Delta =T-t$ and on the observation time scale $\delta 
$ used to construct the volatility process. That this latter dependence
might actually be very weak, seems to be implied by comparison with the data
from several markets.

A closed-form expression for the returns distribution and its asymptotic
behavior may be obtained, namely 
\begin{equation}
P_{\delta }\left( r\left( \Delta \right) \right) =\frac{1}{4\pi \theta
k\delta ^{H-1}\sqrt{\Delta }}\frac{1}{\sqrt{\lambda }}\left. \left( e^{-%
\frac{1}{C}\left( \log \lambda -\frac{d}{dz}\right) ^{2}}\Gamma \left(
z\right) \right) \right| _{z=\frac{1}{2}}  \label{3.10}
\end{equation}
with asymptotic behavior, for large returns 
\begin{equation}
P_{\delta }\left( r\left( \Delta \right) \right) \sim \frac{1}{\sqrt{\Delta
\lambda }}e^{-\frac{1}{C}\log ^{2}\lambda }  \label{3.15}
\end{equation}
with 
\[
r\left( \Delta \right) =\log S_{T}-\log S_{t}\;,\;\theta =e^{\beta
}\;,\;\Delta =T-t\;,\;\lambda =\frac{\left( r\left( \Delta \right)
-r_{0}\right) ^{2}}{2\Delta \theta ^{2}} 
\]
and 
\[
r_{0}=\left( \mu -\frac{\sigma ^{2}}{2}\right) \left( T-t\right)
\;,\;C=8k^{2}\delta ^{2H-2} 
\]

\begin{figure}[tbh]
\begin{center}
\psfig{figure=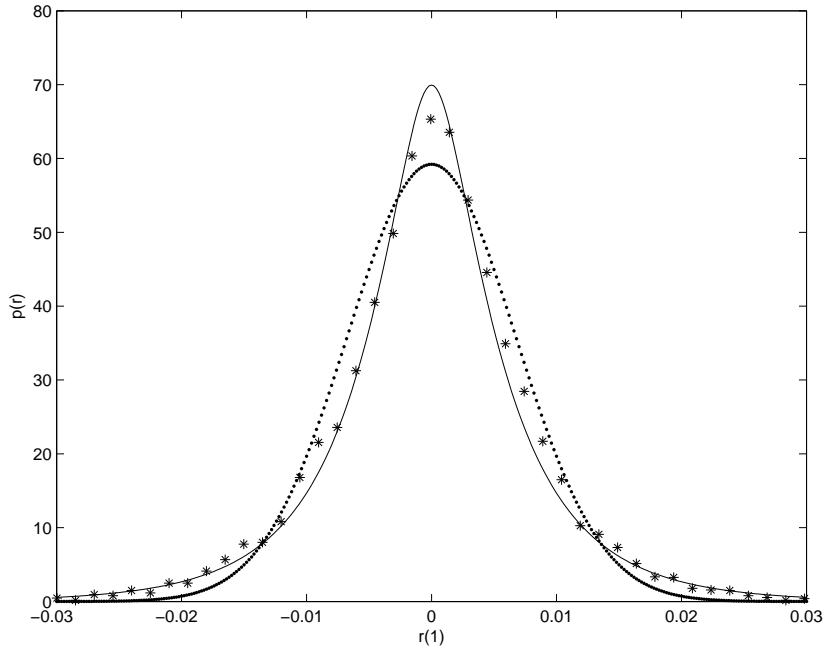,width=11truecm}
\end{center}
\caption{One-day NYSE returns compared with the model predictions and the
lognormal}
\end{figure}

Some illustrative comparisons with market data were performed. In Fig.1
NYSE\ one-day data was used to fix the parameters of the volatility process.
Then, using $H=0.83,$ $k=0.59,$ $\beta =-5,$ $\delta =1$, the one-day return
distribution predicted by the model is compared with the data. The agreement
is quite reasonable. For comparison a log-normal with the same mean and
variance is also plotted in Fig.1. Then, in Fig. 2, using the same
parameters, the same comparison is made for the $\Delta =1$ and $\Delta =10$
data.

\begin{figure}[tbh]
\begin{center}
\psfig{figure=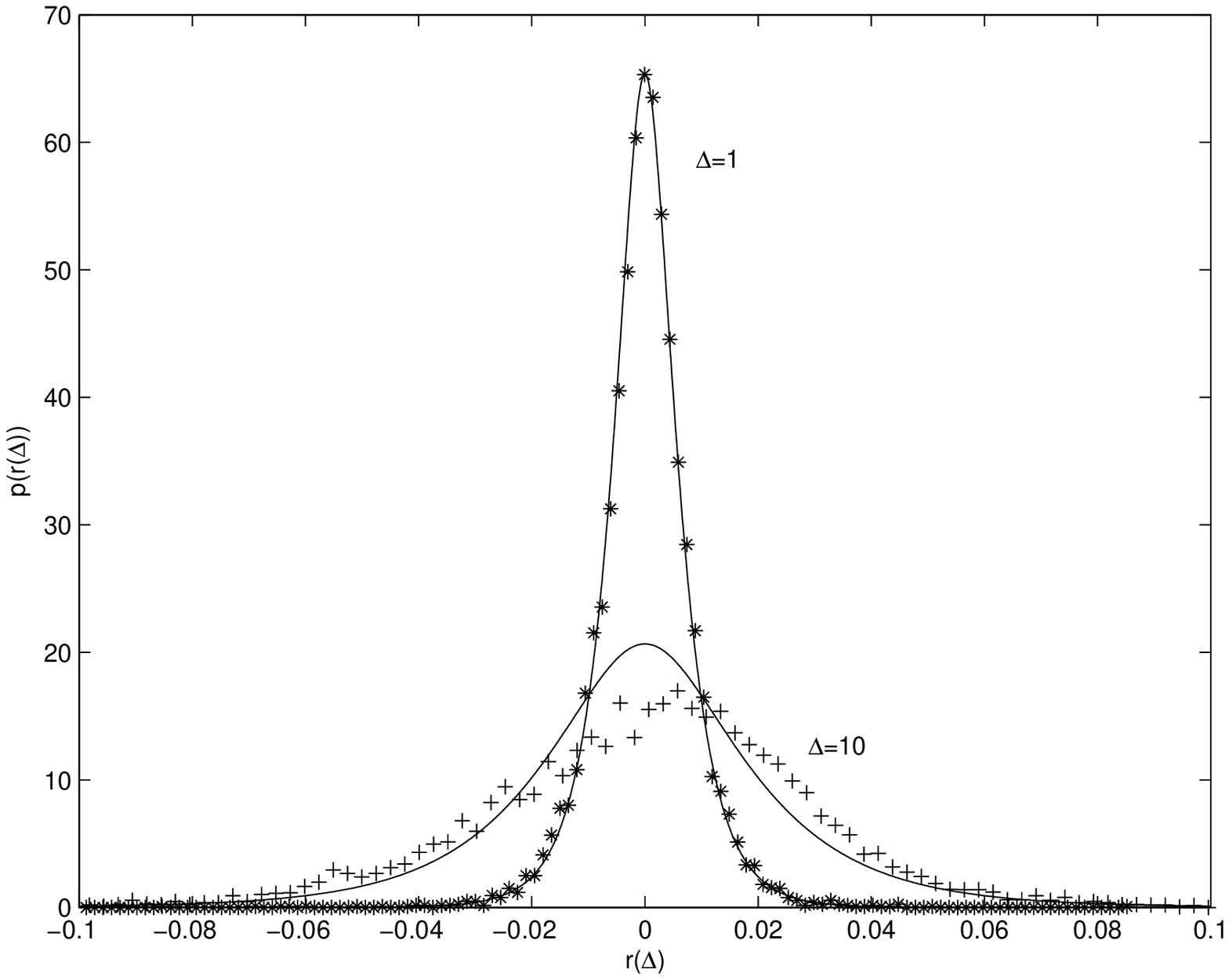,width=11truecm}
\end{center}
\caption{One and ten-days NYSE returns compared with the model predictions}
\end{figure}

\begin{figure}[tbh]
\begin{center}
\psfig{figure=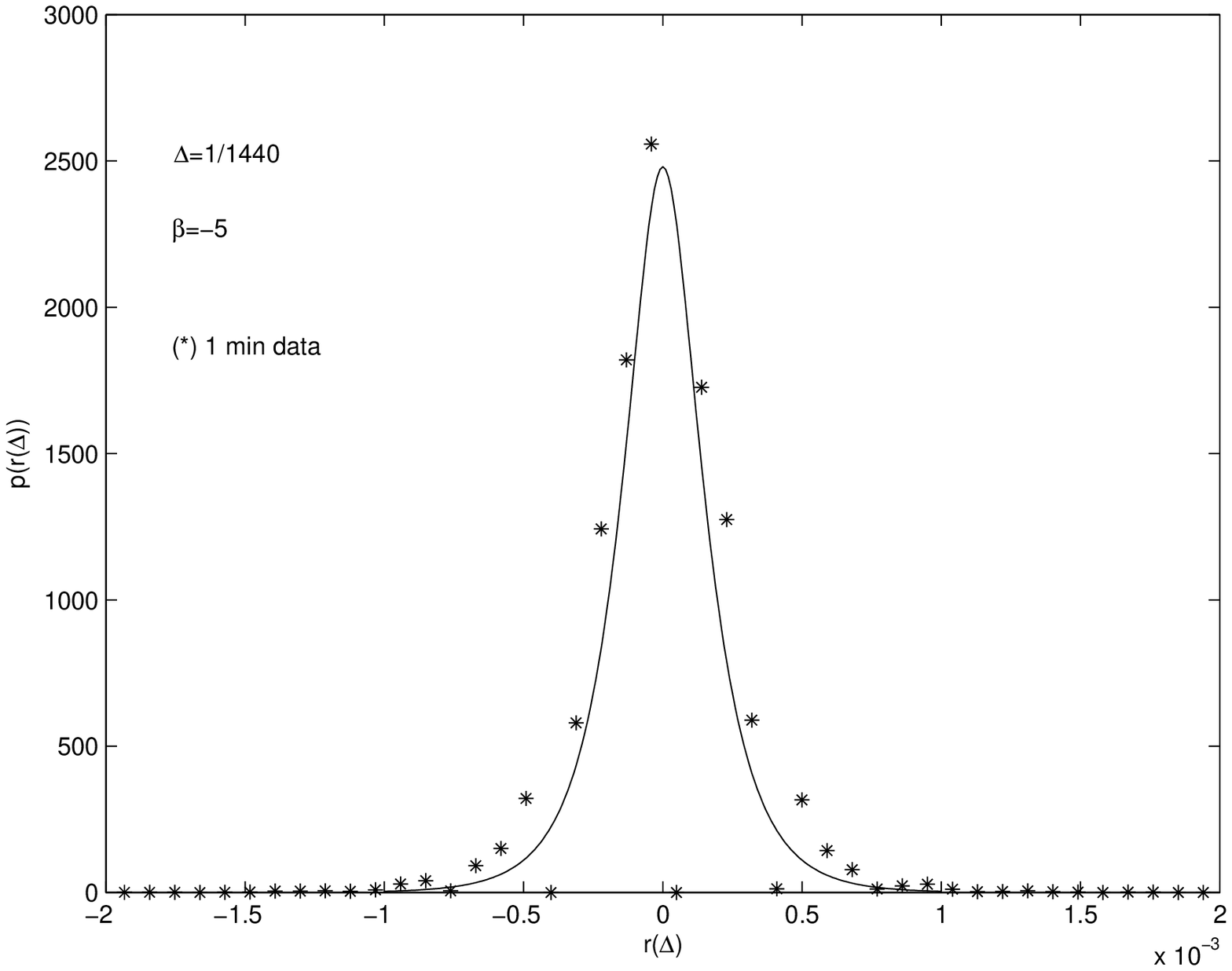,width=11truecm}
\end{center}
\caption{One-minute USD-Euro returns compared with the model predictions,
with parameters obtained from one-day NYSE data}
\end{figure}

Fig. 3 shows a somewhat surprising result. Using the same parameters and
just changing $\Delta $ from $1$ (one day) to $\Delta =\frac{1}{440}$ (one
minute), the prediction of the model is compared with one-minute data of
USDollar-Euro market for a couple of months in 2001. The result is
surprising, because one would not expect the volatility parametrization to
carry over to such a different time scale and also because one is dealing
with different markets.

\begin{figure}[tbh]
\begin{center}
\psfig{figure=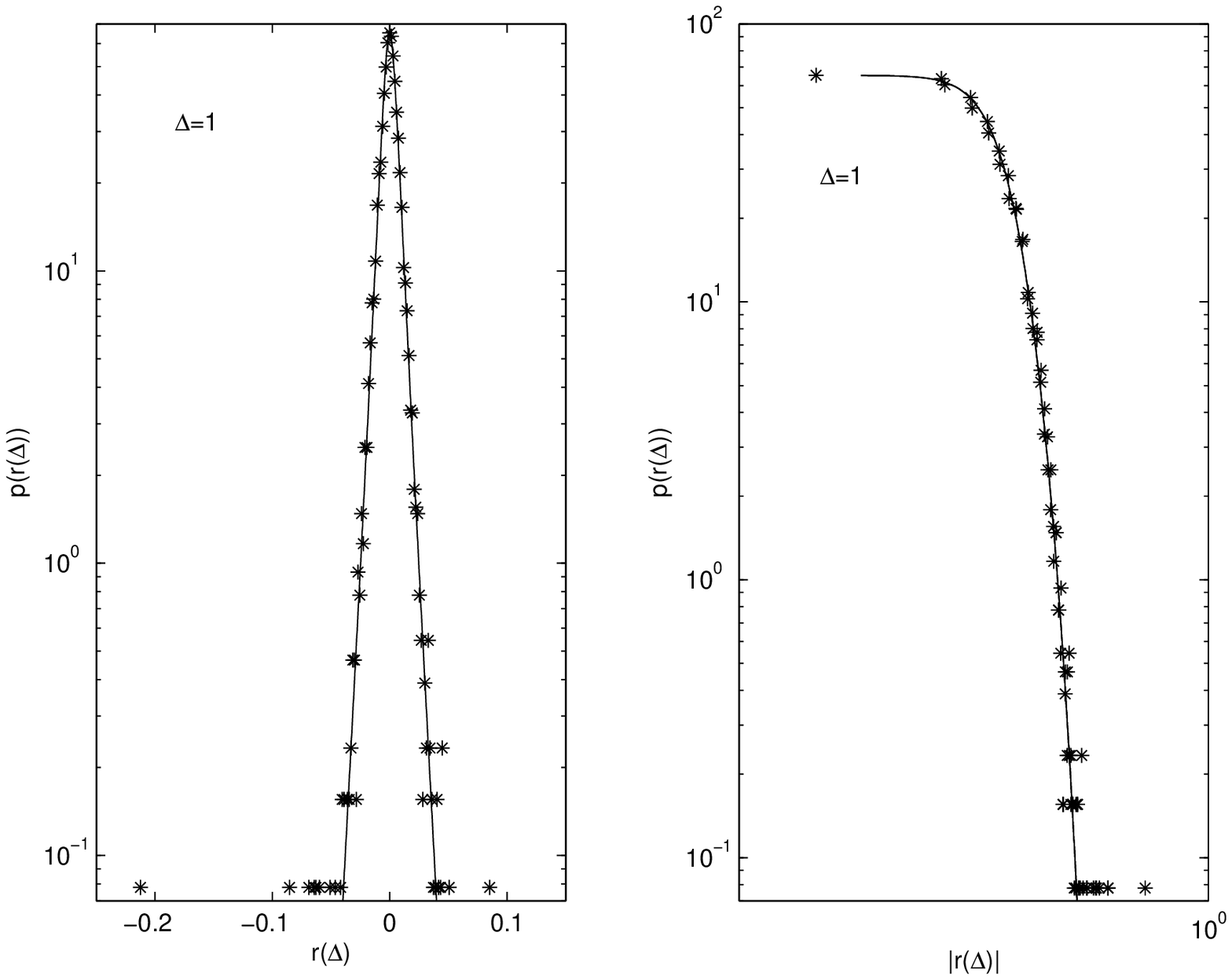,width=11truecm}
\end{center}
\caption{Semilogarithmic and loglog plots of NYSE data}
\end{figure}

\begin{figure}[tbh]
\begin{center}
\psfig{figure=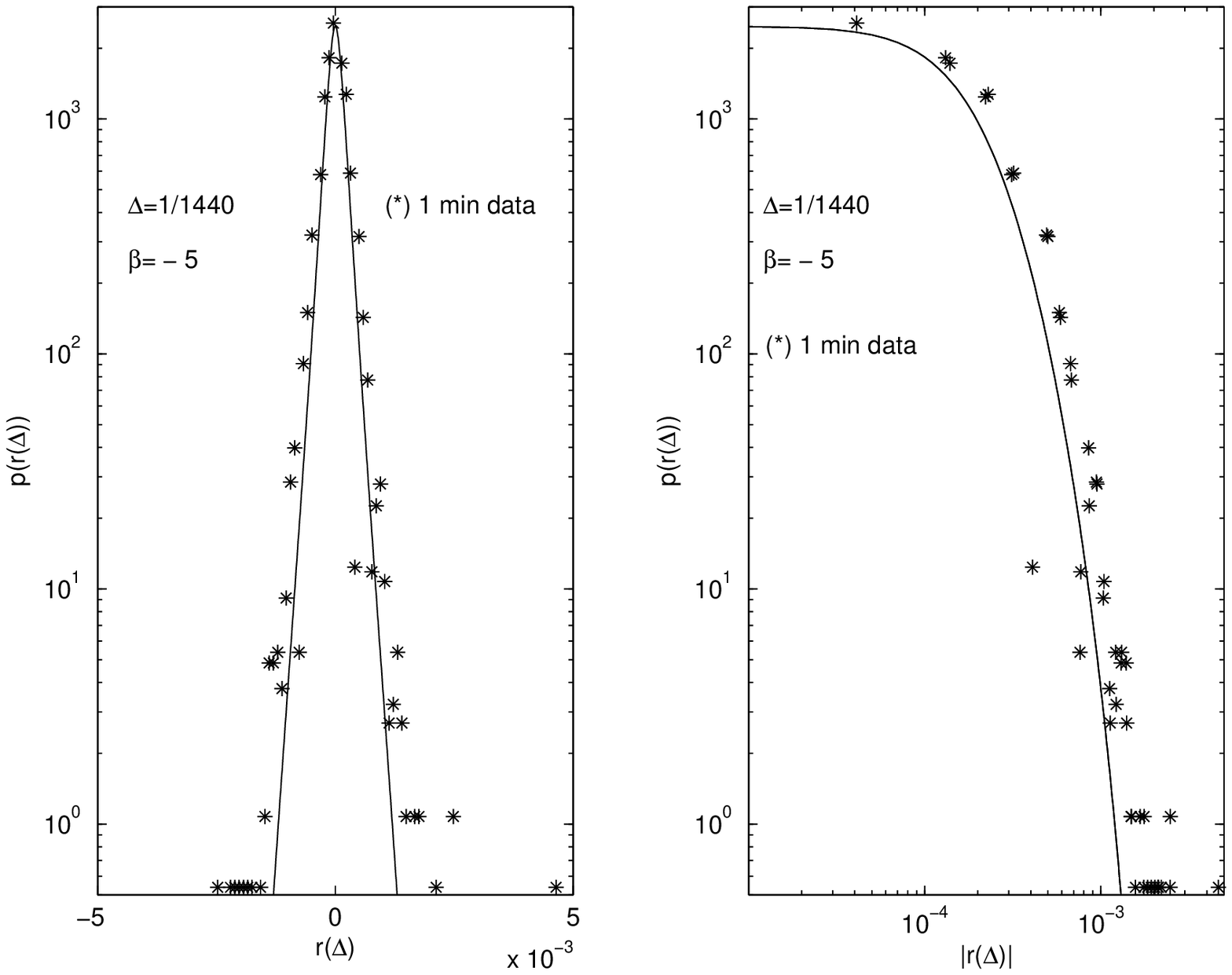,width=11truecm}
\end{center}
\caption{Semilogarithmic and loglog plots of USD-Euro data}
\end{figure}

In Fig.4 and Fig.5 one sees the same one-day and one-minute return data
discussed before, as well as the predictions of the model, both in
semilogarithmic and loglog plots.

As seen from Figs. 4 and 5, the exact result (\ref{3.10}) or (\ref{3.15})
resembles the double exponential distribution recognized by Silva, Prange
and Yakovenko\cite{Silva} as a new stylized fact in market data. The double
exponential distribution has been shown, by Dragulescu and Yakovenko\cite
{Dragulescu}, to follow from Heston's \cite{Heston} stochastic volatility
model. Notice however that our model is different from Heston's model in
that volatility is driven by a process with memory (fractional noise). As a
result, despite the qualitative similarity of behavior at intermediate
return ranges, the analytic form of the distribution and the asymptotic
behavior are different.

\subsection{Option pricing}

New option pricing pricing formulas may be obtained from the model both in a
simplified risk-neutral form or, more accurately, using fractional Malliavin
calculus. Assuming risk neutrality \cite{Cox1}, the value $V\left(
S_{t},\sigma _{t},t\right) $ of an option is the present value of the
expected terminal value discounted at the risk-free rate 
\begin{equation}
V\left( S_{t},\sigma _{t},t\right) =e^{-r\left( T-t\right) }\int V\left(
S_{T},\sigma _{T},T\right) p\left( S_{T}|S_{t},\sigma _{t}\right) dS_{T}
\label{4.1}
\end{equation}
$V\left( S_{T},\sigma _{T},T\right) =\max \left[ 0,S-K\right] $ and the
conditional probability for the terminal price depends on $S_{t}$ and $%
\sigma _{t}$. $K$ is the strike price, $T$ the maturity time and $S_{t}$ and 
$\sigma _{t}$ the price and volatility of the underlying security.

In stochastic volatility models (with or without fractional noise)
risk-neutrality is not an accurate assumption. Nevertheless it provides an
approximate estimate of the deviations from Black-Scholes implied by the
fractional volatility model. As in Hull and White \cite{Hull}, one uses the
relation between conditional probabilities of related variables, namely 
\begin{equation}
p\left( S_{T}|S_{t},\sigma _{t}\right) =\int p\left( S_{T}|S_{t},\overline{%
\log \sigma }\right) p\left( \overline{\log \sigma }|\log \sigma _{t}\right)
d\left( \overline{\log \sigma }\right)  \label{4.2}
\end{equation}
$\overline{\log \sigma }$ being the random variable 
\begin{equation}
\overline{\log \sigma }=\frac{1}{T-t}\int_{t}^{T}\log \sigma _{s}ds
\label{4.3}
\end{equation}
that is, $\overline{\log \sigma }$ is the mean volatility from time $t$ to
the maturity time $T$ conditioned to an average value $\log \sigma _{t}$ at
time $t$. Finally the result for $V\left( S_{t},\sigma _{t},t\right) $ is%
\cite{VilOliv} 
\begin{equation}
V\left( S_{t},\sigma _{t},t\right) =S_{t}\left[ aM\left( \alpha ,a,b\right)
+bM\left( \alpha ,b,a\right) \right] -Ke^{-r\left( T-t\right) }\left[
aM\left( \alpha ,a,-b\right) -bM\left( \alpha ,-b,a\right) \right]
\label{4.16}
\end{equation}
\begin{eqnarray}
M\left( \alpha ,a,b\right) &=&\frac{1}{2\pi \alpha }\int_{-1}^{\infty
}dy\int_{0}^{\infty }dxe^{-\frac{\log ^{2}x}{2\alpha ^{2}}}e^{-\frac{y^{2}}{2%
}\left( ax+\frac{b}{x}\right) ^{2}}  \label{4.17} \\
&=&\frac{1}{4\alpha }\sqrt{\frac{2}{\pi }}\int_{0}^{\infty }dx\frac{e^{-%
\frac{\log ^{2}x}{2\alpha ^{2}}}}{ax+\frac{b}{x}}\mathnormal{\mathnormal{erf}%
\mathnormal{c}}\left( -\frac{ax}{\sqrt{2}}-\frac{b}{\sqrt{2}x}\right) 
\nonumber
\end{eqnarray}
$\mathnormal{erf}c$ is the complementary error function and $a$ and $b$ are 
\begin{equation}
\begin{array}{lll}
a & = & \frac{1}{\sigma _{t}}\left( \frac{\log \frac{S_{t}}{K}}{\sqrt{T-t}}+r%
\sqrt{T-t}\right) \\ 
b & = & \frac{\sigma _{t}}{2}\sqrt{T-t}
\end{array}
\label{4.7}
\end{equation}

\begin{figure}[htb]
\begin{center}
\psfig{figure=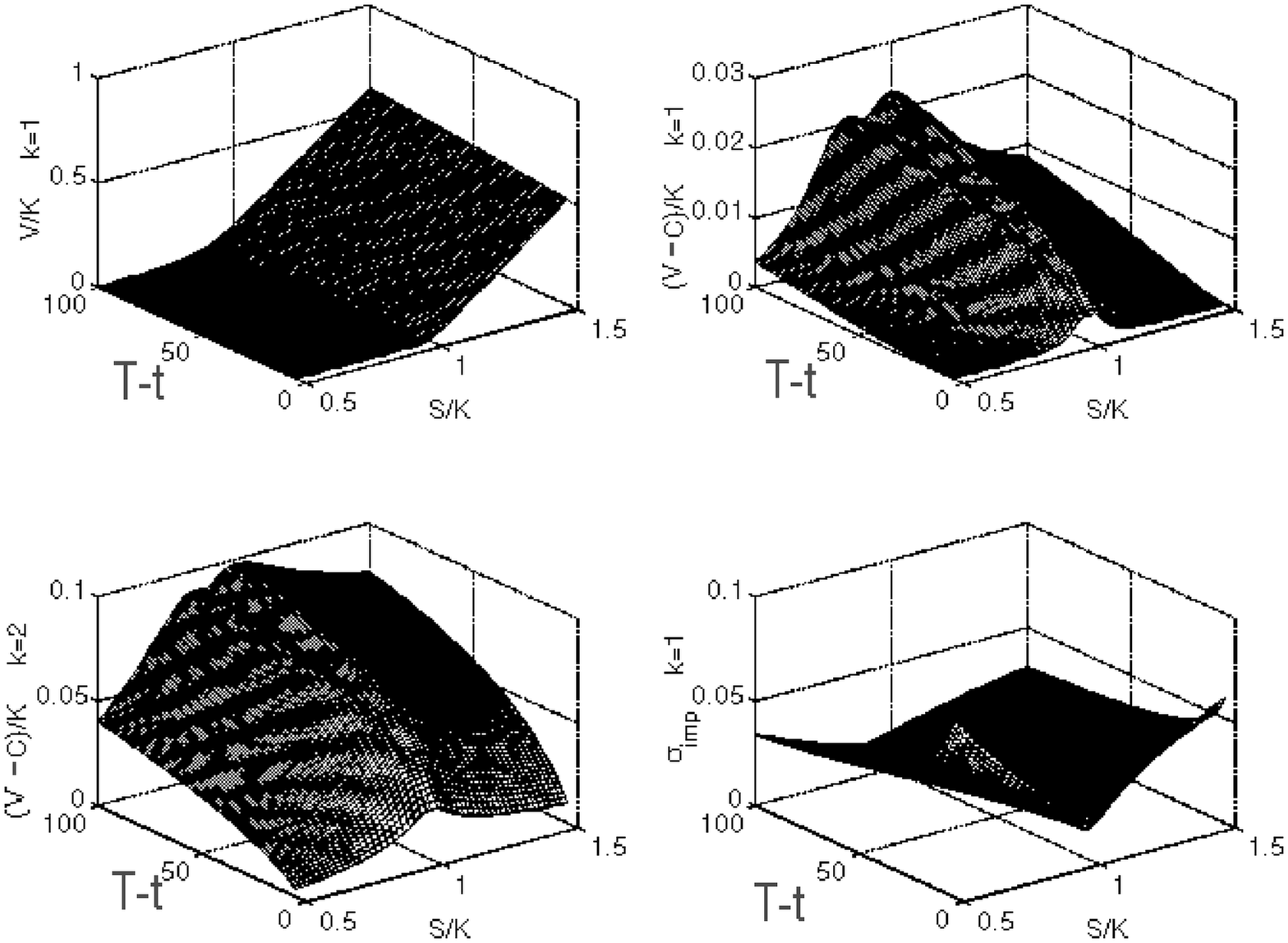,width=11truecm}
\end{center}
\caption{Option price and equivalent implied volatility in the
"risk-neutral" approach to the stochastic volatility model}
\end{figure}

In Fig.6 one plots the option value surface for $V\left( S_{t},\sigma
_{t},t\right) $ in the range $T-t\in [5,100]$ and $S/K\in [0.5,1.5]$ as well
as the difference $\left( V\left( S_{t},\sigma _{t},t\right) -C\left(
S_{t},\sigma _{t},t\right) \right) /K$ for $k=1$ and $k=2$. The other
parameters are fixed at $\sigma =0.01,r=0.001,\delta =1,H=0.8$. To compare
the predictions of the option pricing formula (\ref{4.16}) with the
classical Black-Scholes (BS) result\cite{Black} \cite{Merton}, the implied
volatility required in BS to reproduce the same results was computed. This
is plotted in the lower right panel of Fig.6 which shows the implied
volatility surface corresponding to $V\left( S_{t},\sigma _{t},t\right) $
for $k=1$. One sees that, when compared to BS, it predicts a smile effect
with the smile increasing as maturity approaches.

\subsection{Leverage}

The following nonlinear correlation of the returns 
\begin{equation}
L\left( \tau \right) =\left\langle \left| r\left( t+\tau \right) \right|
^{2}r\left( t\right) \right\rangle -\left\langle \left| r\left( t+\tau
\right) \right| ^{2}\right\rangle \left\langle r\left( t\right) \right\rangle
\label{4.10}
\end{equation}
is called \textit{leverage} and the \textit{leverage effect} is the fact
that, for $\tau >0$, $L\left( \tau \right) $ starts from a negative value
and decays to zero whereas for $\tau <0$ it has almost negligible values.
Fig.7 shows $L\left( \tau \right) $ computed for the NYSE\ index one-day
data in the period 1966-2000.

\begin{figure}[tbh]
\begin{center}
\psfig{figure=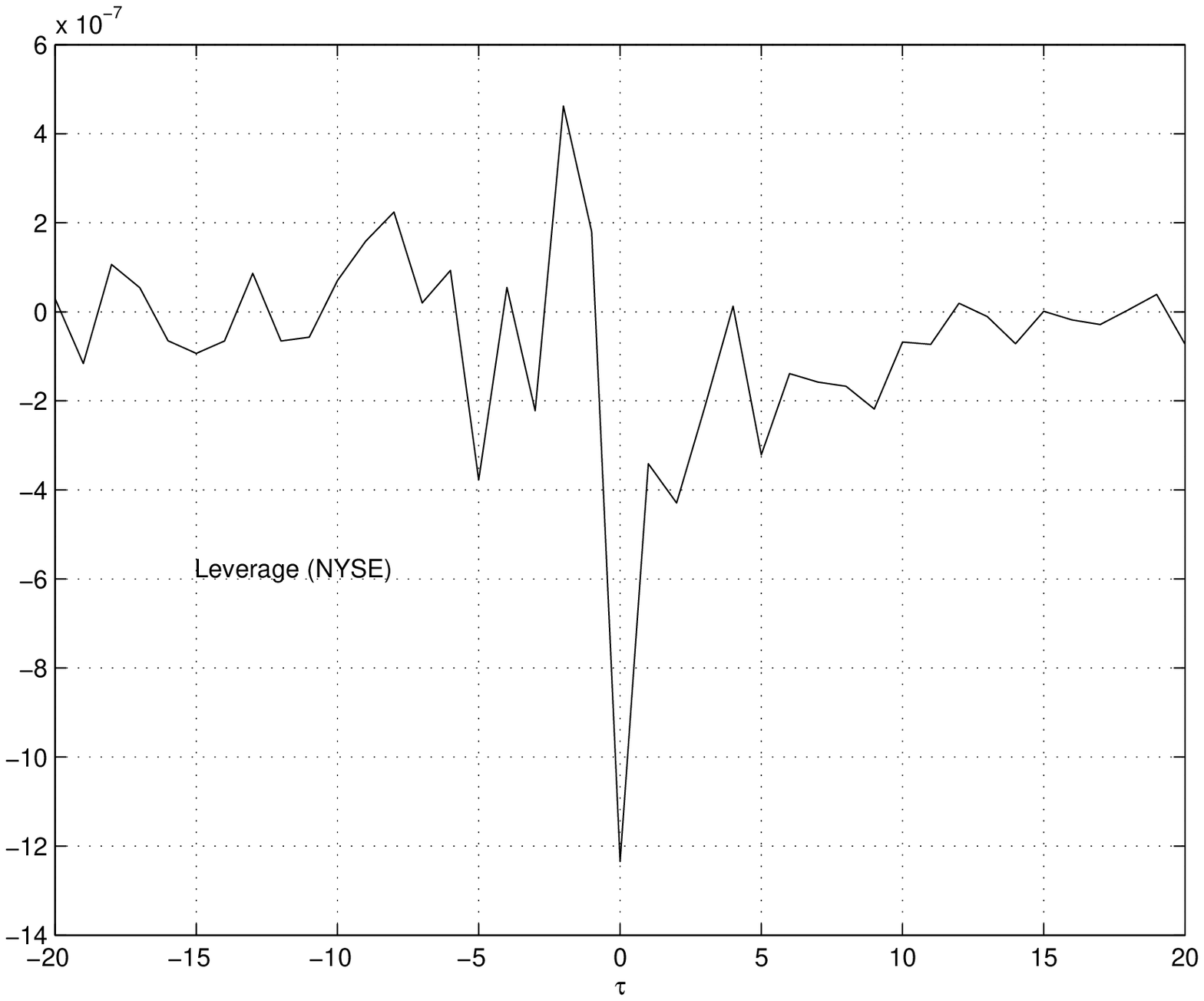,width=11truecm}
\end{center}
\caption{Leverage for the NYSE one-day data in the period 1966-200}
\end{figure}

\begin{figure}[tbh]
\begin{center}
\psfig{figure=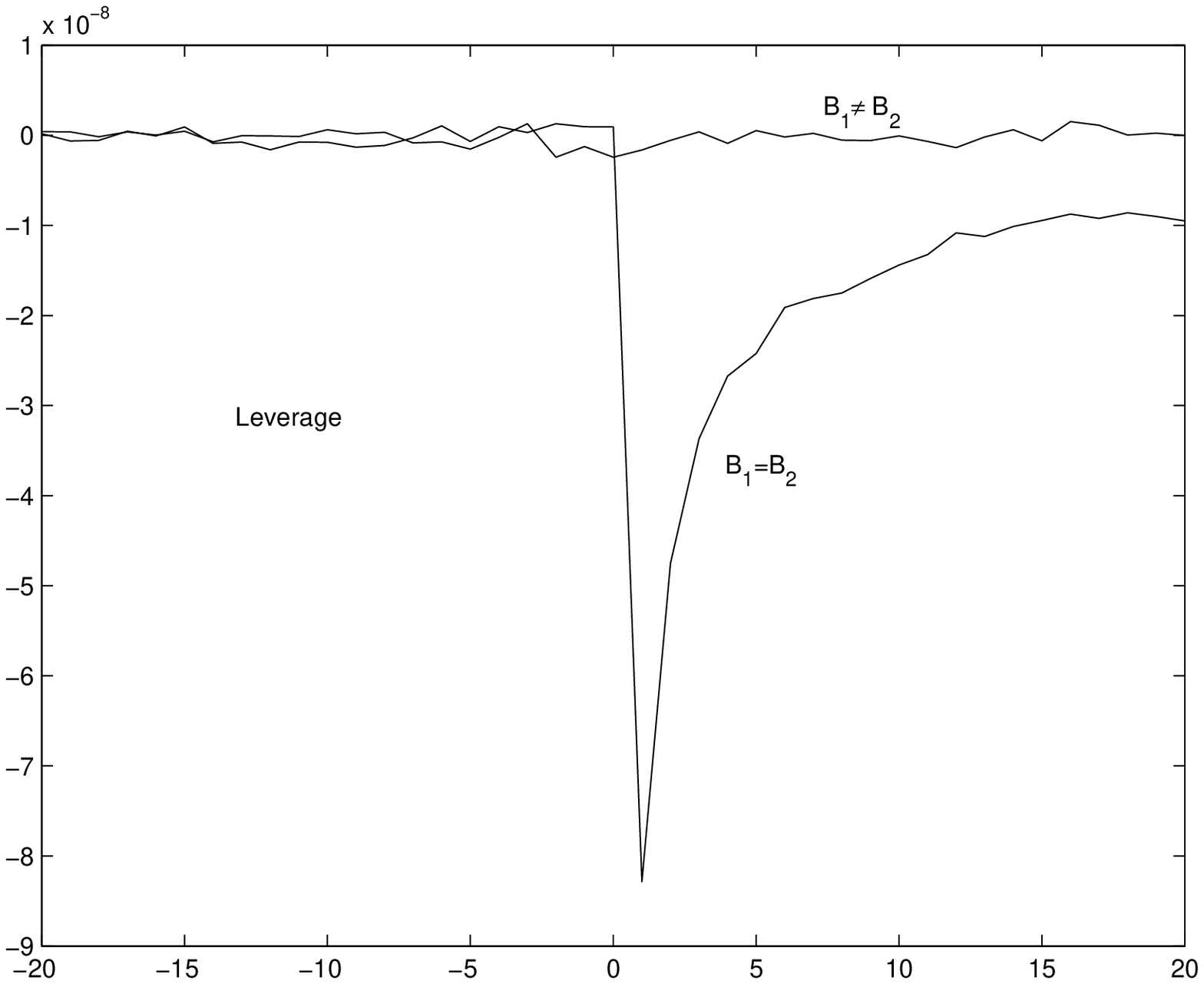,width=11truecm}
\end{center}
\caption{Leverage in the fractional volatility model}
\end{figure}

The leverage behavior of the fractional volatility model will now be
examined. For this purpose it will be convenient to use the following
integral representation of fractional Brownian motion \cite{Embrechts}. 
\begin{equation}
B_{H}\left( t\right) =C\left\{ \int_{-\infty }^{0}\left[ \left( t-s\right)
^{H-\frac{1}{2}}-\left( -s\right) ^{H-\frac{1}{2}}\right] dB\left( s\right)
+\int_{0}^{t}\left( t-s\right) ^{H-\frac{1}{2}}dB\left( s\right) \right\}
\label{4.11}
\end{equation}
Using this representation the fractional volatility model may be rewritten
as 
\begin{equation}
\begin{array}{lll}
dS_{t} & = & \mu S_{t}dt+\sigma _{t}S_{t}dB^{(1)}\left( t\right) \\ 
\log \sigma _{t} & = & \beta +k^{^{\prime }}\int_{-\infty }^{t}\left(
t-s\right) ^{H-\frac{3}{2}}dB^{(2)}\left( s\right)
\end{array}
\label{4.12}
\end{equation}
where $B^{(1)}\left( s\right) $ and $B^{(2)}\left( s\right) $ different
Brownian processes. In Fig.8 one shows the leverage $L\left( \tau \right) $
computed for the model (\ref{4.12}) with $\beta $ and $k^{^{\prime }}$
chosen to match the statistical parameters of the NYSE\ index. Both the $%
B^{(1)}\left( s\right) \neq B^{(2)}\left( s\right) $ and the $B^{(1)}\left(
s\right) =B^{(2)}\left( s\right) $ cases are considered. One sees that for $%
B^{(1)}\left( s\right) \neq B^{(2)}\left( s\right) $ there is no leverage
effect, whereas for $B^{(1)}\left( s\right) =B^{(2)}\left( s\right) $ an
effect is found. Therefore one sees that, identifying the random generator
of the log-price process with the stochastic integrator of the volatility,
at least a part of the leverage effect is taken into account.

\section{Two agent-based studies}

Many factors play a role in a real market. To take into account all the
factors in a model is neither possible nor, in many cases, illuminating. The
objective is to isolate some of the more relevant mechanisms that presumably
play a role in the market and, by stripping the model from other
(inessential?) complications, to exhibit and understand the \textit{purified}
effect of these factors. As in other branches of science, the splitting
apart of the dynamical components of a phenomena, may improve its
understanding \cite{Roehner}.

Two stylized models will be considered. In the first the traders strategies
play a determinant role. In the second the determinant effect is the
limit-order book dynamics, the agents having a random nature.

\subsection{Agent strategies and market impact}

A market model with either random self-adapted strategies or fixed
strategies was studied in detail in \cite{VilStrucgen}. There it was found
that the dominance of two types of strategies was to a large extent
determined by the initial conditions. Different types of return statistics
corresponded to the relative importance of either value investors or
technical traders. The occurrence of market bubbles also corresponded to
situations where technical trader strategies were well represented.

Here, that model will be used for comparison purposes with the fractional
volatility parametrization. The basic ingredients of the model are
summarized below:

One considers a set of investors playing \textit{against }the market, that
is, they have some effect on an existing market that is influenced by other
factors (other investors and general economic effects). This assumption
implies that in addition to the impact of this group of investors on the
market, the other factors are represented by a stochastic process. Therefore 
\begin{equation}
z_{t+1}=f\left( z_{t},\omega _{t}\right) +\eta _{t}  \label{5.10}
\end{equation}
represents the change in the log price $\left( z_{t}=\log p_{t}\right) $
with $\omega _{t}$ being the total investment made by the group of traders
and $\eta _{t}$ the stochastic process that represents all the other
factors. No conservation law is assumed for the total amount of \textit{stock%
} $s$ and \textit{cash} $m$ detained by the group of traders. If $p_{t}$ is
the price of the traded asset at time $t$, the purpose of the investors is
to have an increase, as large as possible, of the total wealth $%
m_{t}+p_{t}\times s_{t}$ at the expense of the rest of the market.

The collective variable is $z$ and each investor payoff at time $t$ is 
\begin{equation}
\Delta _{t}^{(i)}=\left( m_{t}^{(i)}+p_{t}\times s_{t}^{(i)}\right) -\left(
m_{0}^{(i)}+p_{0}\times s_{0}^{(i)}\right)  \label{5.11}
\end{equation}

\textbf{Market impact}

Let $p$ be the price of a representative asset, $z=\log (p)$ and $\omega
_{t} $ the total sum of the buying and selling orders (in money units) for
the asset. Buying orders are positive and selling ones negative. The effect
of these orders on the price change of the asset is called the \textit{%
market impact function}. Let small orders have an impact according to the
loglinear law\cite{Farmer1} \cite{Bouchaud} 
\begin{equation}
z_{t+1}-z_{t}=\frac{\omega _{t}}{\lambda }+\eta _{t}  \label{5.12}
\end{equation}
The constant $\lambda $, is sometimes called the \textit{liquidity}. Eq.(\ref
{5.12}) corresponds naturally to a first order expansion and satisfies the
condition 
\begin{equation}
p(p(p_{0},\omega ^{(1)}),\omega ^{(2)})=p(p_{0},\omega ^{(1)}+\omega ^{(2)})
\label{5.13}
\end{equation}
which is expected to be valid for small orders. However, as pointed out by
Zhang\cite{Zhang3} there is experimental evidence that this is not an
accurate representation for large orders. Therefore a slightly different
market impact function will be used. The reasoning used to motivate it, has
some relation to Zhang's although the result is somewhat different.

When using Eq.(\ref{5.12}) in a discrete-time dynamical model we are somehow
neglecting the fact that the market takes different times to fulfill (and to
react to) small and large orders. Therefore this should be taken into
account when reducing the dynamics to a sequence of equal time steps. In
particular, the reaction of the market may be parametrized by a change in
the $\lambda $ coefficient which, being also related to some stochastic
process, may vary by a factor proportional to $t^{\alpha }$. Taking the time 
$t$ to fill an order to be proportional to its size, one obtains 
\begin{equation}
z_{t+1}-z_{t}=\frac{\omega _{t}}{\lambda _{0}+\lambda _{1}\left| \omega
_{t}\right| ^{\alpha }}+\eta _{t}  \label{5.14}
\end{equation}
This price impact function was first proposed in \cite{VilStrucgen} with $%
\alpha =\frac{1}{2}$. For small orders it recovers the loglinear
approximation and for very large orders (and $\alpha =\frac{1}{2}$) Zhang's
square root law.\medskip

\textbf{The agent strategies}\medskip

In first-order, two main types of informations are taken into account by the
investors, namely the difference (misprice) between price and perceived
value $v_{t}$ 
\begin{equation}
\xi _{t}-z_{t}=\log (v_{t})-\log (p_{t})  \label{5.15}
\end{equation}
and the variation in time of the price (the price trend) 
\begin{equation}
z_{t}-z_{t-1}=\log (p_{t})-\log (p_{t-1})  \label{5.16}
\end{equation}
Consider now a non-decreasing function $f(x)$ such that $f(-\infty )=0$ and $%
f(\infty )=1$. Two useful examples are 
\begin{equation}
\begin{array}{lll}
f_{1}(x) & = & \theta (x) \\ 
f_{2}(x) & = & \frac{1}{1+\exp (-\beta x)}
\end{array}
\label{5.17}
\end{equation}
The information about misprice and price trend is coded on a four-component
vector $\gamma $%
\begin{equation}
\gamma _{t}=\left( 
\begin{array}{c}
f(\xi _{t}-z_{t})f(z_{t}-z_{t-1}) \\ 
f(\xi _{t}-z_{t})\left( 1-f(z_{t}-z_{t-1})\right) \\ 
\left( 1-f(\xi _{t}-z_{t})\right) f(z_{t}-z_{t-1}) \\ 
\left( 1-f(\xi _{t}-z_{t})\right) \left( 1-f(z_{t}-z_{t-1})\right)
\end{array}
\right)  \label{5.18}
\end{equation}
The strategy of each investor is also a four-component vector $\alpha ^{(i)}$
with entries $-1,$ $0,$ or $1$. $-1$ means to sell, $1$ means to buy and $0$
means to do nothing. Hence, at each time, the investment of agent $i$ is $%
\alpha ^{(i)}\cdot \gamma $ . A \textit{fundamental} (value-investing
strategy) that buys when the price is smaller than the value and sells
otherwise would be $\alpha ^{(i)}=\left( 1,1,-1,-1\right) $ and a pure 
\textit{trend-following} strategy would be $\alpha ^{(i)}=\left(
1,-1,1,-1\right) $ . In this setting the total number of possible strategies
is $3^{4}=81$. The strategies will be labelled by numbers 
\begin{equation}
n^{(i)}=\sum_{k=0}^{3}3^{k}\left( \alpha _{k}^{(i)}+1\right)  \label{5.19}
\end{equation}
Therefore the fundamental strategy is strategy no. $72$ and the pure
trend-following one is no. $60$.

An evolution dynamics may be implemented in the model in the following way.
After a number $r$ of time steps, $s$ agents copy the strategy of the $s$
best performers and, at the same time, have some probability to mutate that
strategy. This evolution aims at attaining the goal of improving gains,
while at the same time allowing for some renewal of the strategies. The
percentage of each strategy changes in time and one may find whether some of
them become dominating or stable and when this may occur.

\begin{figure}[tbh]
\begin{center}
\psfig{figure=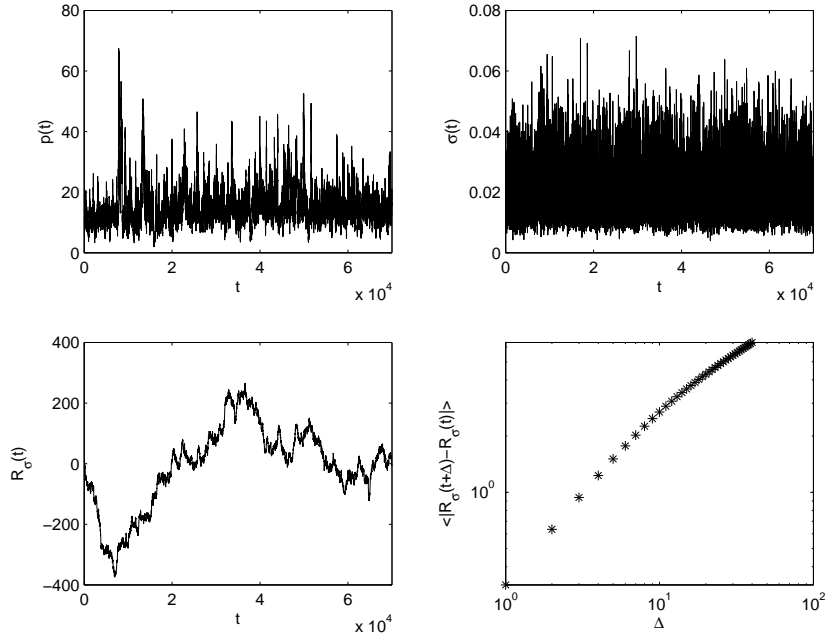,width=11truecm}
\end{center}
\caption{Price, volatility and $R_{\sigma }\left( t\right) $ with equal
amounts of fundamental and trend-following traders}
\end{figure}

The model was run with different initial conditions and with or without
evolution of the strategies. The following results were obtained in \cite
{VilStrucgen}:

- When in the initial condition all traders have the fundamental strategy
and evolution is activated, this strategy stays dominant, not being invaded
by any other of the strategies that are created by the mutation process.
There are however a few other strategies that, after being created, survive
the selection process. This is true for example for the strategies $%
45=\left( 0,1,-1,-1\right) $, $18=\left( -1,1,-1,-1\right) $, $73=\left(
1,1,-1,0\right) $ and $75=\left( 1,1,0,-1\right) $. These surviving
strategies are however similar to the fundamental one. When there is
dominance of the fundamental strategies, the price increments $dp$ have a
Gaussian distribution.

- The fundamental strategy ceases to be stable if it occurs in the initial
condition in smaller amounts $\left( \leq 40\%\right) $.

- For a completely random mixture of strategies in the initial condition,
although the selection mechanism still favors at each evaluation cycle the
best performers, the system never organizes itself to make the total payoff
of this group of traders to grow, nor does a clear dominant strategy emerges.

- When the simulation is run without evolution, with a fixed $50\%$ of
fundamental strategies (no. 72) and $50\%$ of trend-following ones (no. 60),
one sees a large number of bubbles and crashes in the price evolution and
the price increments distribution has fat tails.

Because this last case is the one where the returns statistics is closer to
the actual market data, it is here further analyzed to see whether it also
displays the other features of the fractional volatility model. From typical
simulation runs one computes 
\[
\sigma _{t}^{2}=\frac{1}{\left| T_{0}-T_{1}\right| }\mathnormal{var}\left(
\log p_{t}\right) , 
\]
\[
\sum_{n=0}^{t/\delta }\log \sigma \left( n\delta \right) =\beta t+R_{\sigma
}\left( t\right) 
\]
and 
\[
\left| R_{\sigma }\left( t+\Delta \right) -R_{\sigma }\left( t\right)
\right| 
\]
Fig.9 shows a typical plot of the price process $p\left( t\right) $, the
volatility, $R_{\sigma }\left( t\right) $ and $\mathbb{E}\left\{ \left|
R_{\sigma }\left( t+\Delta \right) -R_{\sigma }\left( t\right) \right|
\right\} $ obtained from the model with equal amounts of \textit{fundamental}
$\left( 1,1,-1,-1\right) $ and \textit{trend-following} $\left(
1,-1,1,-1\right) \,$agents and no evolution. One notices the lack of scaling
behavior of $R_{\sigma }\left( t\right) $ with an asymptotic exponent $0.55$%
, denoting the lack of memory of the volatility process. This might already
be evident from the time behavior of $R_{\sigma }\left( t\right) $ in the
lower left plot. Also, although the returns have fat tails in this case,
they are of different shape from those observed in the market data. Similar
conclusions are obtained with other combinations of agent strategies. In
conclusion: It seems that the features of the fractional volatility model
are not easily captured by a choice of strategies in an agent-based model.
Notice however that what the fractional volatility model parametrizes is the
bulk of the market data, that is, the behavior of the market in normal days.
The agents reactions and strategies are very probably determinant during
market crisis and market bubbles.

\subsection{A limit-order book dynamics model}

Here one considers a limit-order book where \textit{asks} and \textit{bids}
arrive at random on a window $[p-w,p+w]$ around the current price $p$. Every
time a \textit{buy} order arrives it is fulfilled by the closest non-empty
ask slot, the new current price being determined by the value of the ask
that fulfills it. If no ask exists when a buy order arrives it goes to a
cumulative register to wait to be fulfilled. The symmetric process occurs
when a \textit{sell} order arrives, the new price being the bid that buys
it. Because the window around the current price moves up and down, asks and
bids that are too far away from the current price are automatically
eliminated. Sell and buy orders, asks and bids all arrive at random. The
only parameters of the model are the width $w$ of the limit-order book and
the size $n$ of the asks and bids, the sell and buy orders being normalized
to one.

The model was run for different widths $w$ and liquidities $n$ and, for
comparison with the fractional volatility model, one computes as before $%
\sigma _{t}^{2},\sum_{n=0}^{t/\delta }\log \sigma \left( n\delta \right)
=\beta t+R_{\sigma }\left( t\right) $ and $\left| R_{\sigma }\left( t+\Delta
\right) -R_{\sigma }\left( t\right) \right| $.

\begin{figure}[tbh]
\begin{center}
\psfig{figure=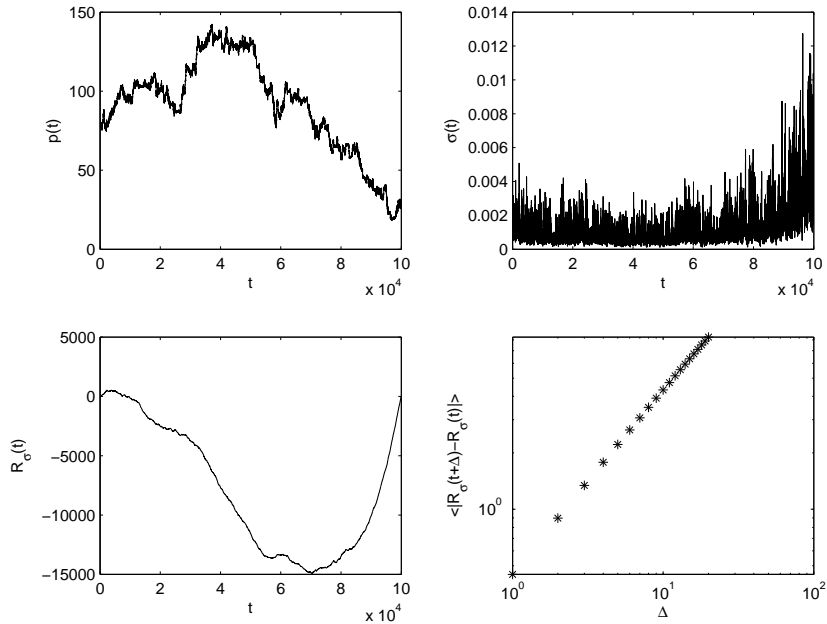,width=11truecm}
\end{center}
\caption{Price, volatility and $R_{\sigma }\left( t\right) $ in the limit
order model}
\end{figure}

\begin{figure}[tbh]
\begin{center}
\psfig{figure=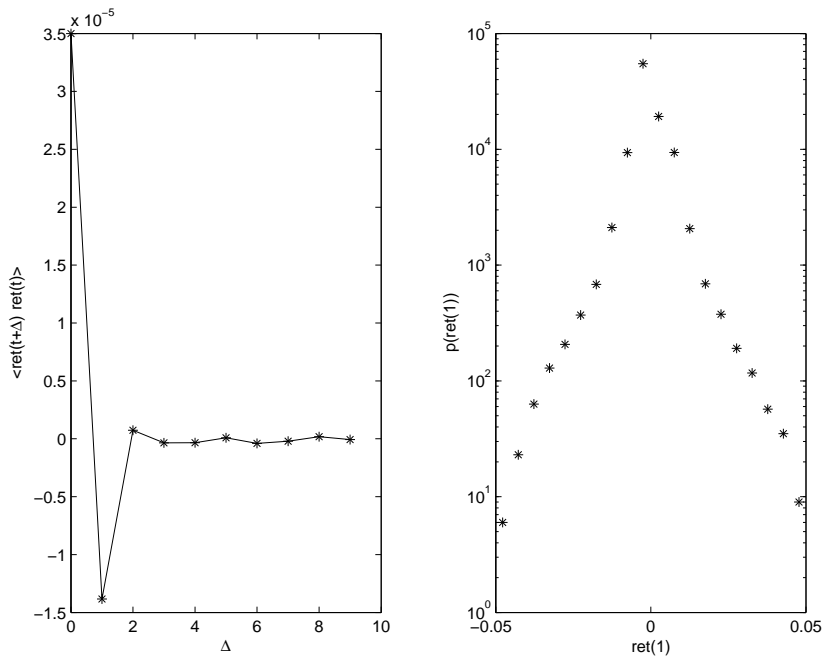,width=11truecm}
\end{center}
\caption{Linear correlation and pdf of returns in the limit order model}
\end{figure}

Although the exact values of the statistical parameters depend on $w$ and $n$%
, the statistical nature of the results seems to be essentially the same.
Fig.10 shows typical plots of the price process $p\left( t\right) $, the
volatility, $R_{\sigma }\left( t\right) $ and $\mathbb{E}\left\{ \left|
R_{\sigma }\left( t+\Delta \right) -R_{\sigma }\left( t\right) \right|
\right\} $ obtained for $n=2$ and the limit-order book divided into $2w+1=21$
discrete price slots with $\Delta p=0.1$. The scaling properties of $%
R_{\sigma }\left( t\right) $ are quite evident from the lower right plot in
the figure, the Hurst coefficient being $0.96$. Fig.11 shows the correlation
and the pdf of the one-time returns. From these results one concludes that
the main statistical properties of the market data (fast decay of the linear
correlation of the returns, non-Gaussianity and volatility memory) are
already generated by the dynamics of the limit-order book with random
behavior of the agents. This implies, as pointed out by some
authors that in the past have considered limit order book models
\cite{Maslov} \cite{Challet1} \cite{Challet2} \cite{Doyne1}, that a 
large part of the market statistical properties (in normal business-as-usual days) 
depends more on the nature of the price fixing financial institutions than on
particular investor strategies.

\section{Conclusions}

(a) The fractional volatility model provides a reasonable mathematical
parametrization of the bulk market data, that is, it captures the behavior
of the market in business-as-usual trading days.

(b) A small modification of the original model, identifying the random
generator of the log-price process and the integrator of the volatility
process, also describes, at least, a part of the leverage effect.

(c) The market statistical behavior in normal days seems to be more
influenced by the nature of the financial institutions (the double auction
process) than by the traders strategies. Specific trader strategies and
psychology should however play a role on market crisis and bubbles.


\begin{thebibliography}{99}
\bibitem{Ding2}  Z. Ding, C. W. J. Granger and R. Engle; \textit{A long
memory property of stock returns and a new model}, Journal of Empirical
Finance 1 (1993) 83-106.

\bibitem{Harvey}  A. C. Harvey; \textit{Long memory in stochastic volatility}%
, Research report 10, London School of Economics, 1993.

\bibitem{Crato}  F. J. Breidt, N. Crato and P. Lima; \textit{The detection
and estimation of long memory in stochastic volatility models,} J. of
Econometrics 83 (1998) 325-348.

\bibitem{Engle}  R. F. Engle; \textit{Autoregressive conditional
heteroscedasticity with estimates of the variance of United Kingdom inflation%
}, Econometrica 50 (1982) 987-1007.

\bibitem{Taylor}  S. J. Taylor; \textit{Modeling stochastic volatility: A
review and comparative study}, Mathematical Finance 4 (1994) 183-204.

\bibitem{Engle2}  R. S. Engle and A. J. Patton; \textit{What good is a
volatility model ?}, Quantitative Finance 1 (2001) 237-245.

\bibitem{VilOliv}  R. Vilela Mendes and M. J. Oliveira; \textit{A
data-reconstructed fractional volatility model}, arXiv:math/0602013

\bibitem{Nualart}  D. Nualart;\ \textit{The Malliavin Calculus and Related
Topics}, Springer-Verlag, Berlin 1995.

\bibitem{Embrechts}  P. Embrechts and M. Maejima; \textit{Selfsimilar
processes}, Princeton Univ. Press, Princeton NJ 2002.

\bibitem{Mandelbrot1}  B. B. Mandelbrot and J. W. Van Ness; \textit{%
Fractional Brownian motions, fractional noises and applications}, SIAM Rev.
10 (1968) 422-437.

\bibitem{Silva}  A. C. Silva, R. E. Prange and V. M. Yakovenko; \textit{%
Exponential distribution of financial returns at mesoscopic time lags: A new
stylized fact}, Physica A344 (2004) 227-235.

\bibitem{Dragulescu}  A. A. Dragulescu and V. M. Yakovenko; \textit{%
Probability distribution of returns in the Heston model with stochastic
volatility}, Quantitative Finance 2 (2002) 443-453.

\bibitem{Heston}  S. L. Heston; \textit{A closed form solution for options
with stochastic volatility with applications to bond and currency options},
The Review of Financial Studies 6 (1993) 327-343.

\bibitem{Cox1}  J. C. Cox and S. A. Ross; \textit{The valuation of options
for alternative stochastic processes}, J. of Financial Economics 3(1976)
145-166.

\bibitem{Hull}  J. C. Hull and A. White; \textit{The pricing of options on
assets with stochastic volatility}, J. of Finance 42 (1987) 281-300.

\bibitem{Black}  F. Black and M. Scholes; \textit{The pricing of options and
corporate liabilities}, J. of Political Economy 81 (1973) 637-654.

\bibitem{Merton}  R. C. Merton; \textit{Theory of rational option pricing},
Bell J. Econ. Manag. Sci. 4 (1973) 141-183.

\bibitem{Roehner}  B. M. Roehner; \textit{The taking apart of economic
phenomena, an essential step,} Econophysics Forum, April - May 1999.

\bibitem{VilStrucgen}  R. Vilela Mendes; \textit{Structure generating
mechanisms in agent-based models}, Physica A295 (2001) 537-561.

\bibitem{Farmer1}  J. D. Farmer; \textit{Market force, Ecology and Evolution;%
} Santa Fe Institute Working Paper 98-12-116.

\bibitem{Bouchaud}  J.-P. Bouchaud and R. Cont; \textit{A Langevin approach
to stock market fluctuations and crashes}, European Phys. Journal B6 (1998)
543.

\bibitem{Zhang3}  Yi-Cheng Zhang; \textit{Toward a theory of marginally
efficient markets}, Physica A 269 (1999) 30.

\bibitem{Maslov} S. Maslov; \textit{Simple model of a limit order-driven market}
arXiv:cond-mat/9910502

\bibitem{Challet1} D. Challet and R. Stinchcombe; \textit{Analyzing and modelling 1+1d markets}
arXiv:cond-mat/0106114

\bibitem{Challet2} D. Challet and R. Stinchcombe; \textit{Limit order market analysis and modelling: on an
universal cause for over-diffusive prices}
arXiv:cond-mat/0211082


\bibitem{Doyne1}  E. Smith, J. Doyne Farmer, L. Gillemot and S.
Krishnamurthy; \textit{Statistical theory of the continuous double auction},
arXiv:cond-mat/0210475
\end{thebibliography}
\end{document}